\documentclass[journal]{IEEEtran}
\usepackage{cite}
\usepackage{amsmath,amssymb,amsfonts}
\usepackage{algorithmic}
\usepackage{graphicx}
\usepackage{textcomp}
\usepackage{xcolor}
\usepackage{comment}

\usepackage{booktabs}
\usepackage{xcolor}
\usepackage{authblk}
\usepackage{multirow}
\usepackage[normalem]{ulem}
\useunder{\uline}{\ul}{}
\newcommand{\Hb}[1]{-\,#1\log #1 - (1-#1)\log(1-#1)}
\newcommand{\ZK}{\frac{1}{K\log 2}}

\ifCLASSINFOpdf
\else
\fi
\usepackage{xcolor}
\newcommand{\framework}{\textsc{DYNAMITE}}
\newcommand{\newframework}{\textsc{SAGE}}

\begin{document}
%
\title{SAGE: Sample-Aware Guarding Engine for Robust Intrusion Detection Against Adversarial Attacks}
%
%
%

\author{Jing~Chen$^{*}$,
        Onat~Gungor$^{*}$,
        Zhengli~Shang,
        and~Tajana~Rosing
\thanks{* Jing Chen and Onat Gungor contributed equally to this work.}
\thanks{Department of Computer Science and Engineering, University of California, San Diego, CA 92093 USA (email: \{jic128, ogungor, z4shang, tajana\}@ucsd.edu).}
}

\maketitle

\begin{abstract}
The rapid proliferation of the Internet of Things (IoT) continues to expose critical security vulnerabilities, necessitating the development of efficient and robust intrusion detection systems (IDS). Machine learning-based intrusion detection systems (ML-IDS) have significantly improved threat detection capabilities; however, they remain highly susceptible to adversarial attacks. While numerous defense mechanisms have been proposed to enhance ML-IDS resilience, a systematic approach for selecting the most effective defense against a specific adversarial attack remains absent. To address this challenge, we previously proposed \framework{}, a dynamic defense selection approach that identifies the most suitable defense against adversarial attacks through an ML-driven selection mechanism. Building on this foundation, we propose \newframework{} (Sample-Aware Guarding Engine), a substantially improved defense algorithm that integrates active learning with targeted data reduction. It employs an active learning mechanism to selectively identify the most informative input samples and their corresponding optimal defense labels, which are then used to train a second-level learner responsible for selecting the most effective defense. This targeted sampling improves computational efficiency, exposes the model to diverse adversarial strategies during training, and enhances robustness, stability, and generalizability. As a result, \newframework{} demonstrates strong predictive performance across multiple intrusion detection datasets, achieving an average F1-score improvement of 201\% over the state-of-the-art defenses. Notably, \newframework{} narrows the performance gap to the Oracle to just 3.8\%, while reducing computational overhead by up to 29$\times$.
\end{abstract}


%
\IEEEpeerreviewmaketitle

\section{Introduction}\label{sec:intro}   
\IEEEPARstart{T}he Internet of Things (IoT) systems connect numerous devices that communicate and share data, enabling smart applications in sectors like healthcare, manufacturing, and transportation \cite{zarpelao2017survey}. IoT systems are particularly susceptible to cyber threats due to their inter-connectivity, resource constraints, and diverse configurations \cite{abiodun2021review}. Consequently, ensuring robust security measures is essential to safeguard these systems against potential attacks. Intrusion Detection Systems (IDS) play a crucial role in identifying and responding to malicious activities within IoT networks by monitoring network traffic and system behavior \cite{zarpelao2017survey}. The integration of machine learning (ML) into IDS has significantly improved their effectiveness in detecting and mitigating cyber threats. ML-IDS possess the capability to analyze vast amounts of data, identify latent patterns, and detect cyberattacks that conventional methods may overlook \cite{da2019internet}. Thus, ML-IDS serve as a robust approach for enhancing IoT security by addressing evolving threats. However, the rise of adversarial attacks poses a significant challenge to the effectiveness of ML-IDS \cite{gungor2024rigorous}. These attacks allow malicious activities to go undetected and harm the security of IoT systems, leading to compromised operations, data breaches, and significant financial losses \cite{mishra2021internet}. 

\begin{figure}[]
\centering
\includegraphics[scale=0.37]{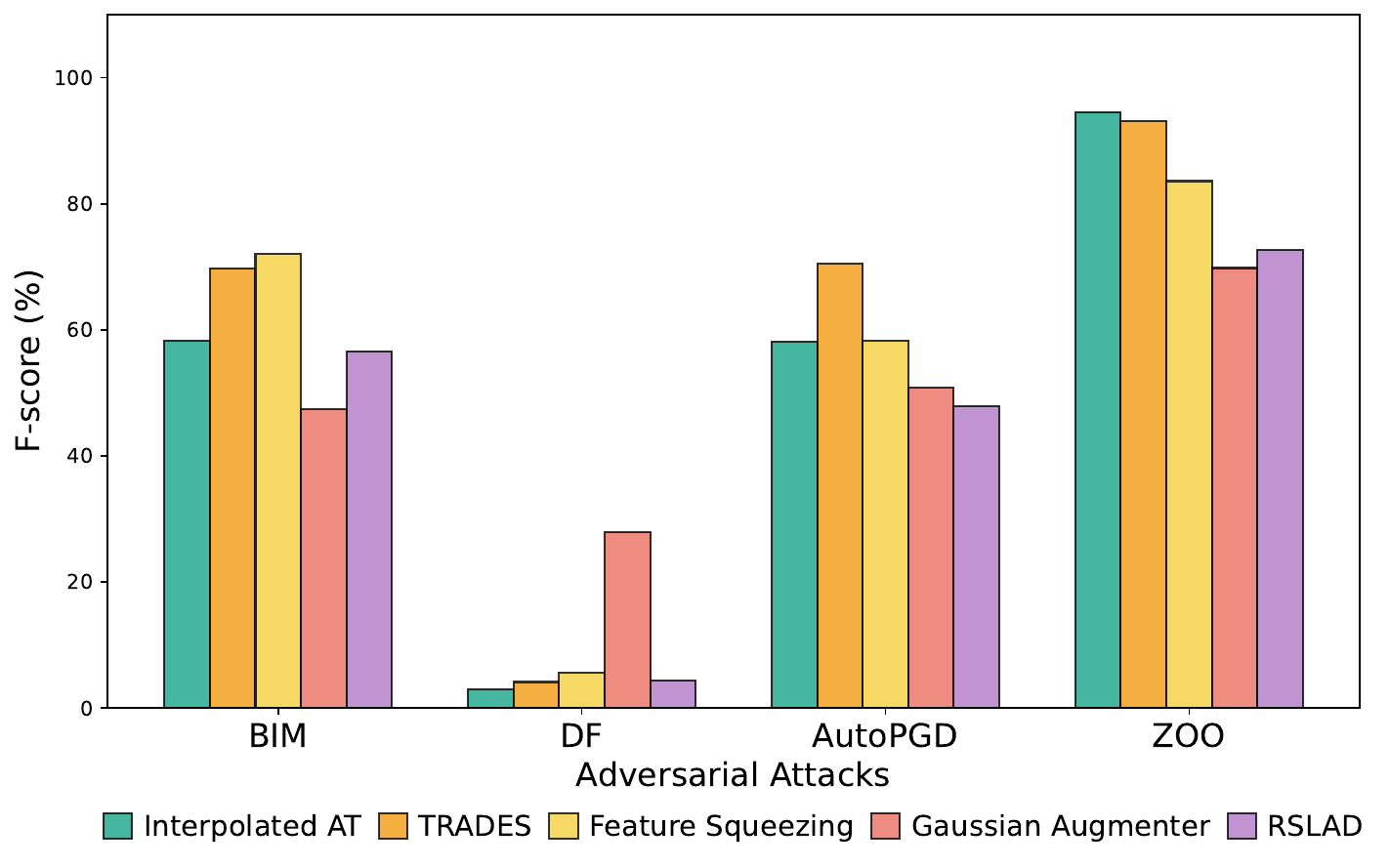}
\caption{State-of-the-art Defense Performance Against Adversarial Attacks}
\label{fig-motivation}
\end{figure}

Developing effective defenses against adversarial attacks is crucial for maintaining the reliability and robustness of ML-IDS \cite{alotaibi2023adversarial}. Several strategies, both general and specific to ML-IDS, have been proposed, including adversarial training \cite{madry2017towards, zhang2019theoretically}, modifications to the training process \cite{papernot2016distillation, zi2021revisiting}, input transformation techniques \cite{xu2017feature}, and methods for adversarial attack detection \cite{debicha2023adv, debicha2023tad}. However, the effectiveness of defense mechanisms varies depending on the specific type of attack they are intended to mitigate \cite{wang2021adversarial}. Given that adversarial attacks can differ in their techniques and objectives, tailored defense strategies are necessary to effectively address each distinct scenario. Fig. \ref{fig-motivation} demonstrates that no single defense model (represented by different colors) is universally effective against all adversarial attacks, with the optimal defense varying depending on the specific nature of the attack (as shown on the x-axis). This variability highlights the limitation of relying on a singular defense mechanism for comprehensive protection. It further emphasizes the importance of a dynamic defense selection mechanism that adaptively assigns the most appropriate defense for each attack scenario. Such an approach is crucial for achieving robust security, as it ensures the real-time deployment of the most effective defense in response to the evolving nature of adversarial attacks.

Building upon our preliminary work, \framework{}~\cite{chen2025dynamite}, this paper addresses two key opportunities for advancing ML-IDS defense frameworks: enhancing computational efficiency and strengthening robustness against previously unseen adversarial attacks. Maximizing efficiency is critical for practical ML-IDS defense deployment in IoT environments, where training on large datasets and running complex defense models pose significant challenges. To this end, our extended framework employs a targeted subsample selection strategy that curates a minimal yet highly informative training set, substantially reducing computational overhead. Furthermore, enhancing robustness against novel threats is essential in an evolving adversarial landscape, where static defenses can be quickly circumvented. To address this, we leverage the Entropic Open-set Active Learning (EOAL) algorithm~\cite{safaei2024entropic}, which is particularly suited to our objectives. Together, these advances yield a defense framework that is both scalable and resilient, systematically enhancing efficiency and robustness.

We introduce the Sample-Aware Guarding Engine (\newframework{}), an adaptive and efficient ML-IDS defense framework. Unlike static defense mechanisms, \newframework{} dynamically selects the optimal defense strategy against adversarial attacks. The \newframework{} pipeline, depicted in Fig.~\ref{framework}, begins with data preprocessing and the training of a baseline ML-IDS model alongside several SOTA defense models. To simulate diverse threat landscapes, we generate adversarial samples using various attack strategies and intensities. Each defense model is then evaluated against these samples, and performance metrics are used to label each sample with its most effective defense. To optimize the training of our defense selection mechanism, \newframework{} employs Entropic Open-set Active Learning (EOAL)~\cite{safaei2024entropic} to identify and prioritize the most informative and diverse samples. This approach significantly reduces the training data volume while preserving high performance. A second-level learner is then trained on this reduced dataset to predict the most effective defense for previously unseen adversarial attacks. Our experiments on multiple realistic intrusion detection datasets show that \newframework{} surpasses state-of-the-art defenses, yielding an average F1-score improvement of 201\%. Moreover, \newframework{} remains robust against previously unseen adversarial attacks, demonstrating its ability to adapt to new threats. Notably, by leveraging EOAL's focus on sparsely represented regions, \newframework{} achieves comparable or superior robustness using only 1\% of labeled data, with gains of up to 3.4 points over full supervision. \newframework{} also substantially improves computational efficiency, achieving up to a 29$\times$ speedup compared to the Oracle, while maintaining only a 3.8\% gap in F1-score. These findings position \newframework{} as a robust and efficient defense framework that achieves high accuracy while substantially reducing computational overhead.


\section{Background and Related Work}\label{sec:related}
\subsection{ML-based Intrusion Detection Systems (ML-IDS)}
The growing dependence on computer networks and the expansion of the Internet of Things (IoT) have introduced significant security challenges, driven by the increasing complexity and diversity of these interconnected systems~\cite{Sebestyen2025Literature}. Intrusion Detection Systems (IDS) are designed to monitor network activity and detect malicious behavior. IDS methods are broadly categorized into two types: signature-based and anomaly-based~\cite{gungor2024rigorous}. Signature-based IDS rely on predefined signatures of known attacks to identify malicious activities, offering high accuracy for detecting known threats but struggling with zero-day attacks. In contrast, anomaly-based IDS detect deviations from normal network behavior, enabling them to identify previously unseen attacks, though they may suffer from higher false-positive rates. This limitation has motivated the integration of ML techniques into IDS, enhancing their ability to identify complex and evolving attack patterns with greater accuracy~\cite{gungor2024roldef}. A range of ML models, such as Decision Trees (DT), Random Forests (RF), and Deep Neural Networks (DNN), have been utilized in ML-IDS to improve detection capabilities~\cite{liu2019machine}. By leveraging large datasets, these models can learn intricate patterns of network behavior, surpassing the performance of traditional methods. However, despite their effectiveness, ML-IDS solutions remain vulnerable to adversarial attacks, where malicious actors manipulate input data to evade detection or degrade system performance~\cite{gungor2024rigorous}.

\subsection{Adversarial Attacks}
Adversarial attacks manipulate ML models by introducing small, intentional, and often imperceptible perturbations to input data~\cite{gungor2022stewart}. These attacks are especially critical in ML-IDS since they can exploit model vulnerabilities to evade detection~\cite{gungor2024rigorous}. While white-box attacks have full access to the model’s details for gradient-based perturbations, black-box attacks generate adversarial examples without internal knowledge, using queries or transfer methods \cite{papernot2017practical}. Overall, we select nine state-of-the-art, widely used white-box and black-box adversarial attacks: Fast Gradient Sign Method (FGSM)~\cite{goodfellow2014explaining}, Basic Iterative Method (BIM)~\cite{kurakin2018adversarial}, Projected Gradient Descent (PGD)~\cite{madry2017towards}, Auto Projected Gradient Descent (AutoPGD)~\cite{croce2020reliable}, DeepFool (DF)~\cite{moosavi2016deepfool}, Zeroth Order Optimization (ZOO) \cite{chen2017zoo}, Scale-Invariant Nesterov Iterative Fast Gradient Sign Method (SINI-FGSM)~\cite{lin2020Nesterov}, Variance-tuned Nesterov Iterative Fast Gradient Sign Method (VNI-FGSM)~\cite{wang2021Enhancing}, and Cost-aware Feasible Attack (CaFA)~\cite{bentov2025CaFA}.

\subsection{Adversarial Defenses}
Defense mechanisms aim to protect ML models from adversarial attacks by enhancing their robustness or reducing the effectiveness of adversarial perturbations \cite{goyal2023survey}. We evaluate ten representative defenses across four principal categories: (i) \emph{Adversarial Training (AT)}: methods that incorporate adversarial examples during training to improve model robustness, including Projected Gradient Descent AT (PGD-AT)~\cite{madry2017towards}, Interpolated AT (IAT)~\cite{lamb2019interpolated}, TRadeoff-inspired Adversarial DEfense via Surrogate-loss Minimization (TRADES)~\cite{zhang2019theoretically}, and Free AT (FAT)~\cite{shafahi2019adversarial}. (ii) \emph{Training-Process Modification}: approaches that enhance robustness via training-time augmentation or architectural adjustments without explicit inner maximization, exemplified by Gaussian Augmenter (GA)~\cite{zantedeschi2017efficient} and Robust Generative Adaptation Network (RGAN)~\cite{li2025robust}. (iii) \emph{Robust Network Design}: architectural or knowledge-distillation-based methods that smooth decision boundaries and improve stability, including Defensive Distillation (DD)~\cite{papernot2016distillation} and Robust Soft Label Adversarial Distillation (RSLAD)~\cite{zi2021revisiting}. (iv) \emph{Input Preprocessing}: inference-time transformations applied to inputs to mitigate adversarial perturbations, represented by Feature Squeezing (FS)~\cite{xu2017feature} and Gaussian Noise (GN)~\cite{kassam2012signal}. These defenses form the static portfolio from which the selector assigns the most appropriate defense to each sample.

\subsection{Adversarial Defense Mechanisms in ML-IDS}
Several efforts have been directed toward developing adversarial defense mechanisms specifically tailored to enhance the robustness of ML-IDS against adversarial attacks. Han et al. \cite{han2021evaluating} address traffic-space attacks targeting ML-based NIDS, proposing a defense scheme that reduces evasion rates across multiple attack scenarios. Debicha et al. \cite{debicha2023adv} introduce Adv-Bot, a framework for generating adversarial botnet traffic to test and strengthen IDS defenses. Debicha et al. \cite{debicha2023tad} present a transfer learning-based framework that employs multiple adversarial detectors to improve detection rates. Alslman et al.~\cite{alslman2025breaking} enhance ML-IDS resilience in 5G networks with a CycleGAN-based adversarial recovery mechanism. Similarly, Holla et al.~\cite{holla2025adversarial} propose a dual-layered defense for ML-IDS in cloud environments, which combines adversarial training with SHAP-based feature selection. Existing studies on ML-based IDS often rely on isolated or manually selected defense mechanisms, which limits their generalizability. These methods also remain highly sensitive to subtle adversarial perturbations, leading to undetected intrusions in dynamic environments.


In contrast, \newframework{} frames defense as a per-sample decision over the state-of-the-art methods. Rather than assuming adversarial attack knowledge, it trains a second-level learner to select the most suitable defense. The resulting attack-agnostic mechanism integrates evidence across defenses during training and, at deployment, assigns an appropriate defense to each sample in real time. This shift from static defenses to data-driven, sample-aware selection improves adaptability in evolving threat environments while keeping inference lightweight and deployment practical for ML-IDS.

\subsection{Dynamic Defense Methods}
Dynamic defense strategies have been investigated across several cybersecurity domains. Zheng et al.~\cite{ZHENG2022Dynamic} examined approaches such as Moving Target Defense and Mimic Defense to increase attacker cost through surface reconfiguration and redundancy. El Gadal and Ganti~\cite{Gadal2024Dynamic} proposed a reinforcement learning–based framework for dynamically selecting intrusion detection and mitigation techniques in SDN environments. Rehman et al.~\cite{REHMAN2024Proactive} introduced a proactive mechanism that combines OS diversity and cyber deception to deter attackers in IoT networks. Feng et al.~\cite{FENG2024Security} developed a deep reinforcement learning algorithm to adaptively allocate resources under adversarial conditions. To the best of our knowledge, no prior work has investigated dynamic defense for ML-based intrusion detection systems under adversarial attacks. Existing approaches, while effective in specific domains, often depend on predefined policies or simplified attacker assumptions, limiting their applicability in evolving adversarial settings. In contrast, we frame dynamic defense selection as a data-driven, attack-agnostic process that enables per-sample adaptation to both conventional cyber threats and adversarial manipulations.

\subsection{Active Learning}
Active learning (AL) aims to reduce labeling costs by selectively querying the most informative samples from a large unlabeled pool under a fixed budget~\cite{tharwat2023survey}. This is especially valuable when annotation is costly, class distributions are imbalanced, or the data-generating process evolves, as it enables models to refine decision boundaries while covering rare or underrepresented cases. Commonly used AL strategies include uncertainty sampling (selecting points the model is least confident about), Query-by-Committee (selecting samples where multiple models disagree), density-weighted methods (prioritizing uncertain samples in dense regions of the data), and core-set or diversity-based selection (choosing batches that broadly cover the feature space). These strategies are model- and domain-agnostic and are frequently used as baselines for evaluating new active learning techniques~\cite{ren2021survey}.

AL has also been applied to domain-specific tasks. For example, Vaarandi and Guerra-Manzanares \cite{VAARANDI2024Network} use AL to prioritize NIDS alerts, reducing labeling costs. Alaa and Schenck \cite{Alaa2025Active} handle missing data by selecting samples based on imputation uncertainty. Fan et al. \cite{FAN2024Integrating} combine AL with semi-supervised learning for fault diagnosis. Despite these advances, active learning has not been widely integrated with dynamic defense selection in ML-IDS, where the learner must choose among defenses per sample. Consequently, empirical evidence for this specific setting remains limited.

\begin{figure*}[]
      \centering
      \includegraphics[scale=0.6]{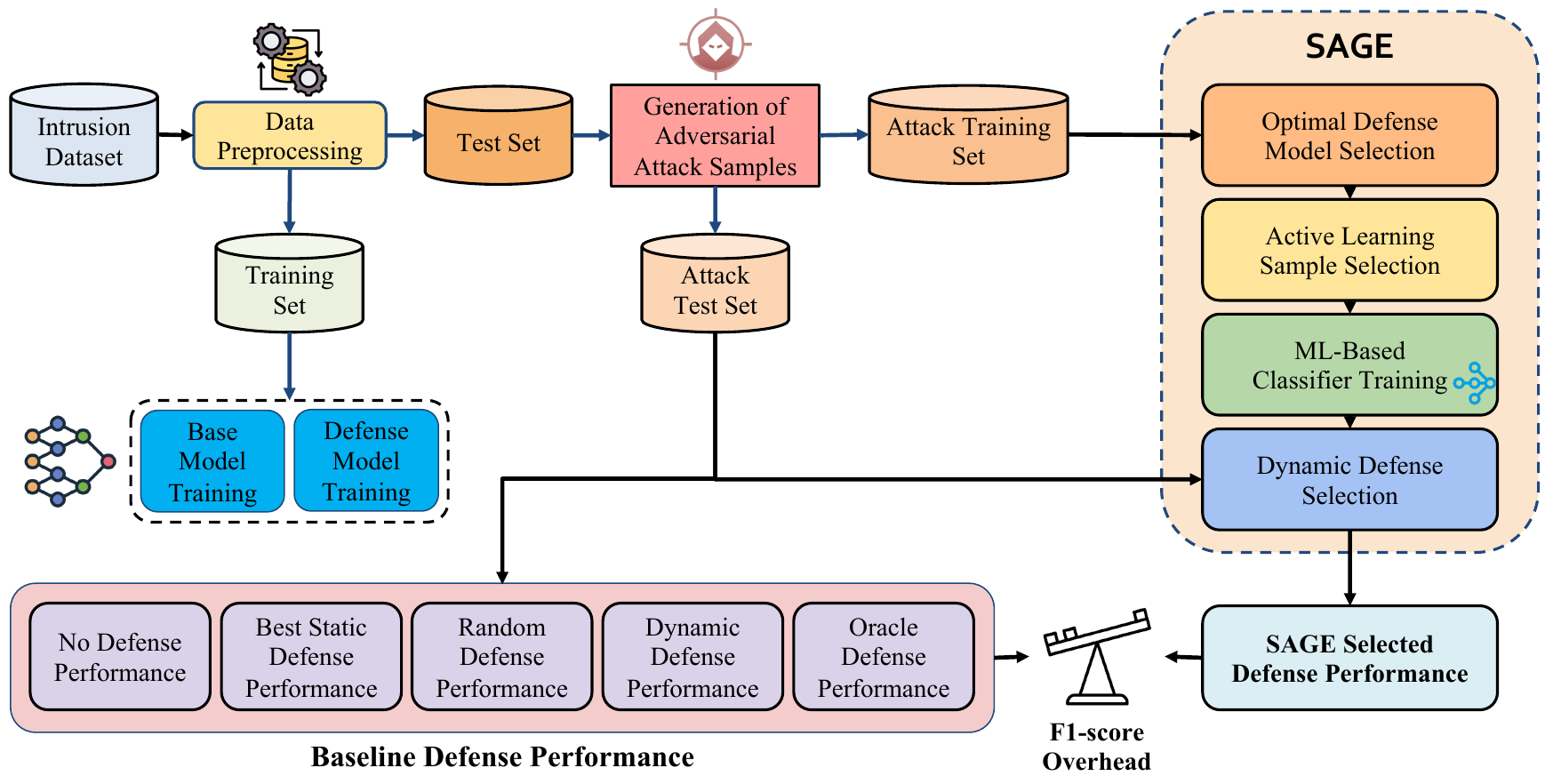}
      \caption{Overview of the Sample-Aware Guarding Engine (\newframework{}) pipeline. Raw data is preprocessed and used to train a baseline ML-IDS model alongside multiple specialized defense models. Adversarial samples are generated across diverse attack strategies and intensities, and each defense model is evaluated to assign optimal defense labels. Entropic Open-set Active Learning (EOAL) identifies the most informative and diverse samples, which are then used to train a second-level learner that dynamically selects the most effective defense for each incoming adversarial sample, ensuring robust and efficient protection.}
      \label{framework}
\end{figure*}

\section{\newframework{} Framework}\label{sec:framework}

We design Sample-Aware Guarding Engine (\newframework{}), a dynamic defense selection framework designed to enhance the robustness of machine learning-based intrusion detection systems against adversarial attacks. Extending \framework{}~\cite{chen2025dynamite}, \newframework{} achieves both superior computational efficiency and enhanced robustness by leveraging targeted subsample selection and a multi-level, active learning–guided per-sample defense selection strategy. \newframework{} integrates several key components—adversarial sample generation, defense model training, and an active learning–guided dynamic defense selection algorithm—forming a comprehensive pipeline that evaluates and mitigates adversarial threats both effectively and efficiently.

As shown in Figure~\ref{framework}, the pipeline begins with data preprocessing, where raw data is cleaned, normalized, and encoded. The preprocessed data is then used to train a baseline DNN model along with multiple adversarial defense models. To simulate real-world scenarios, adversarial datasets are generated using diverse attack strategies, providing a comprehensive benchmark for evaluating framework effectiveness. Each defense model is subsequently assessed to determine its performance across different adversarial scenarios, identifying the most effective strategies for each case. To enable dynamic defense selection, a multi-level learner examines dataset patterns to predict the most effective defense for unseen adversarial samples. This learner combines machine learning with active learning in a multi-layer framework to accurately select the optimal defense model for each input. By integrating optimal defense labels derived from the performance matrix—which evaluates the effectiveness of each defense model against various adversarial attacks—the \newframework{} dynamically selects the most suitable defense for each attack scenario. Finally, the \newframework{}’s performance is compared to that of Oracle, the best static defense models (state-of-the-art defenses), random defense selection, and recommendation-based dynamic selection. In the following subsections, we describe each \newframework{} module in sequence, beginning with data preprocessing, followed by adversarial attack generation, defense training and best defense labeling, the dynamic defense selection algorithm, and concluding with the inference evaluation protocol.

\subsection{Data Preprocessing}
This module performs data cleaning to remove redundant or irrelevant features, standardization to normalize numerical features for consistent scaling, and categorical encoding to convert classification features into numerical representations suitable for ML models. After preprocessing, the data is split into training and test sets. The training set is used to train the baseline and defense models, while the test set is reserved for generating adversarial attacks and conducting final evaluations.


\subsection{Generation of Adversarial Attack Samples}
\subsubsection{Selected Adversarial Attacks} We select nine state-of-the-art adversarial attacks (BIM~\cite{kurakin2018adversarial}, FGSM~\cite{goodfellow2014explaining}, PGD~\cite{madry2017towards}, DF~\cite{moosavi2016deepfool}, AutoPGD~\cite{croce2020reliable}, ZOO~\cite{chen2017zoo}, SINIFGSM~\cite{lin2020Nesterov}, VNIFGSM~\cite{wang2021Enhancing}, and CaFA~\cite{bentov2025CaFA}) to generate adversarial samples. Perturbation magnitude is controlled via the epsilon ($\varepsilon$) parameter, allowing tests under a range of attack intensities. These attacks with varying perturbation magnitudes provide a comprehensive benchmark for assessing the robustness of \newframework{} across diverse adversarial scenarios. 

\subsubsection{Adversarial Dataset Generation}
The generation process involves applying each attack model to the dataset, with $\varepsilon$ values adjusted to simulate varying levels of adversarial intensity. A unique adversarial dataset is generated for each combination of nine attack methods and four epsilon values \{0.01, 0.1, 0.2, 0.3\}, resulting in a total of 36 distinct datasets. Each attack is applied to the test dataset, maintaining the same sample size as the original. This ensures consistent evaluation while introducing adversarial perturbations based on attack type and intensity. After generating adversarial attack samples, we split them into two sets: \textit{attack training} and \textit{attack test}. The training portion is used to train our dynamic defense selection model, while the test portion is used for final evaluation.  

\subsection{Baseline Model Training}
To establish a performance baseline, a Deep Neural Network (DNN)~\cite{al2021x} is trained on the original, unperturbed dataset. The model is then evaluated under different adversarial attack configurations, providing a reference for assessing the effectiveness of defense strategies. This baseline serves as a crucial benchmark, illustrating the impact of adversarial attacks on model performance and emphasizing the importance of robust defense mechanisms and dynamic selection approaches.


\subsection{Defense Model Training}

We leverage ten state-of-the-art defenses against adversarial attacks: Projected Gradient Descent Adversarial Training \cite{madry2017towards}, Interpolated Adversarial Training \cite{lamb2019interpolated}, Tradeoff-inspired Adversarial Defense via Surrogate-loss Minimization (TRADES) \cite{zhang2019theoretically}, Free Adversarial Training \cite{shafahi2019adversarial}, Gaussian Augmenter \cite{zantedeschi2017efficient}, Defensive Distillation \cite{papernot2016distillation}, Robust Soft Label Adversarial Distillation (RSLAD) \cite{zi2021revisiting}, Feature Squeezing \cite{xu2017feature}, Gaussian Noise \cite{kassam2012signal}, and Robust Generative Adaptation Network (RGAN)~\cite{li2025robust}. To address varying defense requirements, we introduce multiple parameter configurations for certain defense models. For RSLAD, configurations like RSLAD10 and RSLAD100 adjust optimization strength to evaluate robustness trade-offs. This approach systematically assesses adaptability to different adversarial perturbation levels. Applying these defense methods to diverse adversarial attacks enables the framework to evaluate model adaptability and performance across attack scenarios. These defenses form the basis of the dynamic selection mechanism, allowing the framework to deploy the most effective strategy for each adversarial sample.

\subsection{Optimal Defense Identification}

\subsubsection{Constructing Attack Training and Attack Test Data}
The attack training and attack test data are created using a subset of the 36 adversarial datasets—generated using different attack methods and epsilon values from the test set—ensuring a distinction between known and unknown data during model evaluation. Specifically, the datasets with an epsilon value of 0.1 (9 datasets) are used as attack training data, representing the known data. The remaining datasets, with remaining epsilon values (27 datasets), serve as attack test data, representing the unknown data. Beyond varying only the perturbation strength, we further evaluate robustness under a more challenging scenario: excluding certain adversarial attack types entirely from the training phase and introducing them only during testing. This extension allows us to assess how well our method generalizes to previously unseen adversarial threats, and we introduce this setup in detail in Section~\ref{defense-selection-setup}.

\subsubsection{Optimal Defense Selection}
To assess the defense models, we process attack training data through all ten defenses and record key metrics, such as the macro F1-score. This generates a ``performance matrix'', where each entry represents a defense model's effectiveness against a specific adversarial sample. The matrix serves as a basis for comparing defenses and identifying best strategies, offering insights into how each model addresses adversarial perturbations. To determine the most effective defense for each adversarial sample, we analyze the performance metrics of all ten defense models and select the highest-performing defense for each sample. This selected defense is then used as the label, which forms the ground truth for training our dynamic defense selection mechanism. 


\subsection{Dynamic Defense Selection Algorithm}
\label{Dynamic_Defense_Selection_Algorithm}

\begin{figure}[]
\centering
\includegraphics[scale=0.33]{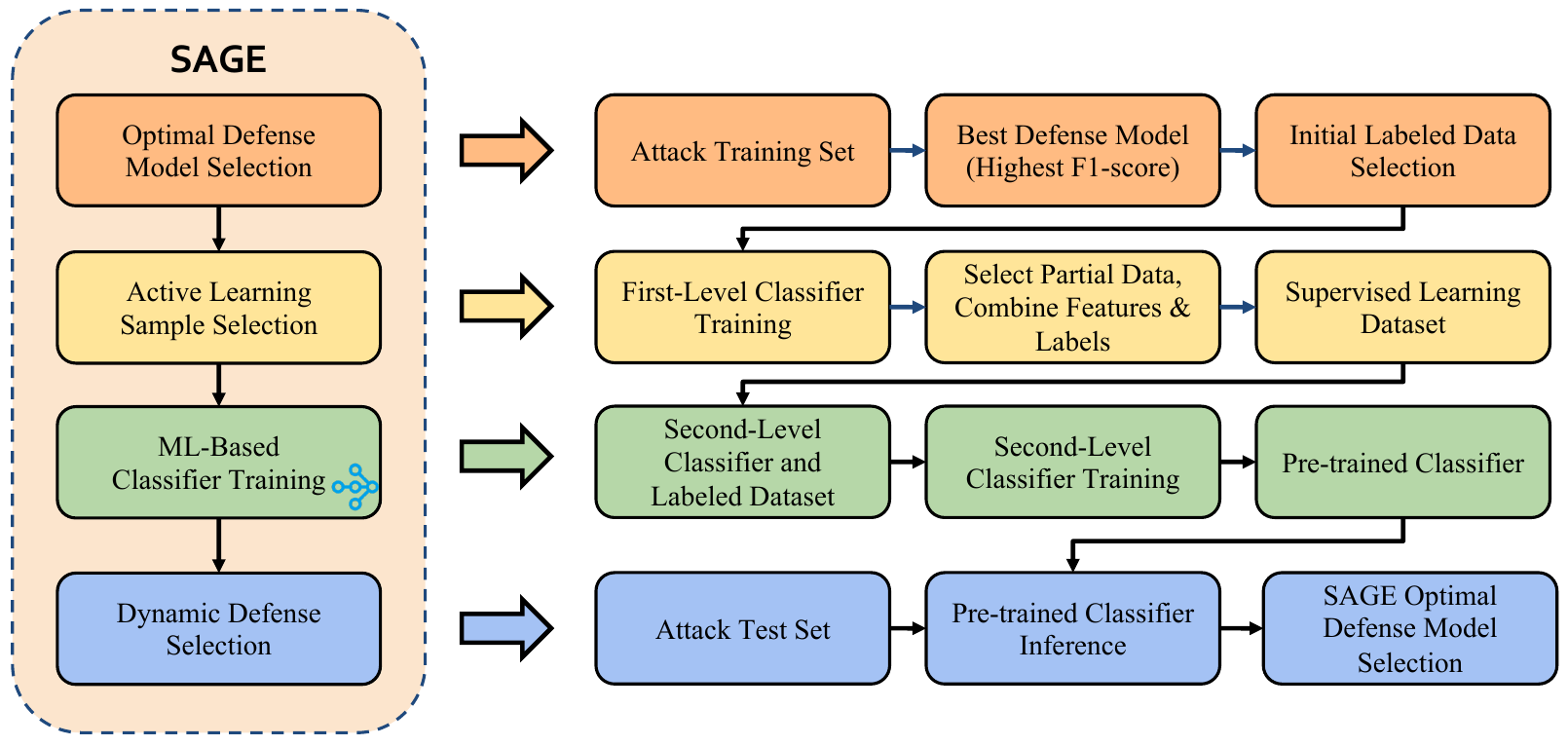}
\caption{Dynamic Defense Selection Algortihm}
\label{EOAL_figures}
\end{figure}

To achieve robust and efficient defense allocation, \newframework{} incorporates a two-layer dynamic defense selection algorithm (Figure~\ref{EOAL_figures}). In the first layer, active learning is employed to identify the most informative samples while minimizing labeling costs. Specifically, we adopt Entropic Open-set Active Learning (EOAL)~\cite{safaei2024entropic} to prioritize uncertain or underrepresented inputs, thereby improving the handling of previously unseen adversarial attacks. In the second layer, a lightweight ensemble-based classifier maps adversarial inputs to their corresponding optimal defenses, enabling efficient and accurate per-sample defense selection.

\subsubsection{Active Learning Sample Selection (First Layer)}
\paragraph{Motivation and Challenges}
A central challenge in our dynamic defense selection approach is training the second-level learner to assign the most effective defenses to adversarial samples. Obtaining the optimal defense labels for this training is \emph{computationally expensive}, as it requires evaluating each defense against multiple attack variants. Active learning offers a potential remedy by selecting and labeling the most informative samples; however, standard active learning methods, while effective at reducing labeling effort, typically assume a fixed input distribution~\cite{cacciarelli2024active}. This assumption breaks down in realistic deployments, where \emph{an adversary can introduce previously unseen attacks by varying attack strength or creating entirely new attack types.} This limitation can lead to a dynamic defense policy that overfits to the training distribution and performs poorly on novel adversarial inputs, highlighting the need for a sample selection strategy that explicitly accounts for distributional shifts.

\paragraph{EOAL for Dynamic Defense Selection}
To overcome the limitations of standard active learning, \newframework{} leverages Entropic Open-set Active Learning (EOAL)~\cite{safaei2024entropic}, which explicitly targets samples that are both uncertain and sparsely represented. In our context, these ``unknowns'' arise in two complementary ways: (i) \emph{variation in attack strength}, where training covers only a single perturbation level of a given attack type but deployment may encounter a broader range, and (ii) \emph{unseen attack types}, resulting from the exclusion of certain attack models during training. To handle (i), we include the same attacks at varying perturbation levels during evaluation, while for (ii) we introduce entirely new attack types at test time. EOAL addresses both types of ``unknowns'' by selecting high-uncertainty samples near decision boundaries affected by perturbation-strength shifts and within sparsely represented regions corresponding to missing attack models~\cite{safaei2024entropic}. By prioritizing these inputs, the second-level learner develops a defense selection policy that improves generalization to previously unseen attacks while minimizing labeling cost, supporting our framework’s goals of scalable and robust ML-IDS defense against adversarial attacks.

\paragraph{EOAL Workflow}
The EOAL process begins with the attack training set and their per-sample optimal defense labels derived from the ``performance matrix''. From this set, 10\% of the data is selected as the initial labeled dataset using \textit{stratified random sampling} (i.e., randomly sampling within each defense class) to ensure balanced representation of attack patterns.
This step corresponds to the first-level classifier training in Fig.~\ref{EOAL_figures} (yellow boxes). We train a lightweight classifier (implemented as a random forest in our experiments) to provide a preliminary mapping from inputs to defense labels, which is used exclusively to guide the selection of unlabeled samples for querying. By combining EOAL with the first-level random forest classifier, we iteratively apply the active learning process to the remaining unlabeled pool over multiple acquisition rounds, gradually increasing the size of the labeled dataset in each round (see Section~\ref{sec:ablation} for an ablation study). Specifically, we rank the unlabeled samples based on the entropy predicted by the random forest and select a diverse batch to minimize redundancy and enable efficient batch processing. Each selected batch is paired with its optimal-defense label to form supervised subsets of different sizes, which are then fed to the second-level learner.

\paragraph{EOAL Methodology and Mathematical Formulation}
EOAL selects samples that provide the most useful information for training the second-level defense selector. For each unlabeled input \(x\), EOAL computes an acquisition score that balances: (i) uncertainty with respect to known defense labels, and (ii) uncertainty arising from potentially unseen (open-set) patterns. Let \(F(\cdot)\) denote the first-level feature extractor, and let the first-level classifier produce one-vs-rest posteriors \(\{p_i(x)\}_{i=1}^K\) over the \(K\) defense labels.

$\triangleright$ \textbf{Closed-set entropy:} This term measures how uncertain the model is about the known defense labels. High uncertainty indicates that the sample lies near the decision boundary of the current classifier. It is computed as the normalized average of one-vs-rest binary entropies:
\begin{align}
\label{eq:Sc}
S_c(x) &= \frac{1}{K} \sum_{i=1}^K H(p_i), \quad p_i := p_i(x)
\end{align}
where \(S_c(x)\in[0,1]\), $H(p_i)$ denotes the binary entropy function.

$\triangleright$ \textbf{Distance-based (open-set) entropy:} To account for unseen or underrepresented attack patterns, EOAL estimates uncertainty based on the sample’s proximity to clusters in feature space that represent open-set regions. Let \(\{c_i\}_{i=1}^K\) be the cluster centers obtained from a working set of unlabeled samples. A soft assignment with temperature \(T>0\) gives
\begin{align}
\label{eq:qi}
q_i(x)
&= \frac{\exp\!\big(-\lVert F(x)-c_i\rVert/T\big)}
         {\sum_{j=1}^K \exp\!\big(-\lVert F(x)-c_j\rVert/T\big)}, \\
\label{eq:Sd}
S_d(x)
&= -\frac{1}{\log K}\sum_{i=1}^K q_i(x)\,\log q_i(x),
\end{align}
where \(S_d(x)\in[0,1]\) measures how scattered the sample is across open-set regions. High values indicate the sample is near sparsely represented regions, which could correspond to unseen attacks.

$\triangleright$ \textbf{Acquisition score:} The final score combines closed-set and open-set uncertainties:
\begin{equation}
\label{eq:score}
S(x) = S_c(x) - S_d(x)
\end{equation}
Samples with the largest \(S(x)\) (high closed-set uncertainty but low open-set proximity) are selected for labeling. This ensures that EOAL focuses on inputs that are informative for the second-level learner, avoiding hard-but-uninformative outliers.

\noindent In our framework, the first-level classifier outputs \(p_i(x)\) and features \(F(x)\). We implement clustering with \(k\)-means on \(F(x)\) to obtain \(K\) centers \(\{c_i\}_{i=1}^K\). EOAL operates in batches; to reduce redundancy, we apply a \emph{farthest-first} diversity filter in feature space: starting from the top-scoring candidate, we iteratively add the sample that maximizes its minimum distance to the current batch until the batch quota is reached. This ensures that selected queries are informative.

\subsubsection{ML-Based Classifier Training (Second Layer)}
The second layer of our dynamic defense selection algorithm trains an ML-based classifier to assign the most suitable defense model to each incoming sample, ensuring robust performance across varying attack conditions. We adopt models that balance resilience to heterogeneous attack patterns with the capacity to model nonlinear feature–defense relationships. Their computational efficiency enables low-latency per-sample inference, making them well-suited for dynamic defense selection.

During training, adversarial samples selected through EOAL are paired with their corresponding optimal defense labels to train the second-level learner. This process enables the model to learn the mapping between input features and effective defense strategies, thereby identifying patterns that inform per-sample defense allocation. At inference time, the trained second-level learner predicts the optimal defense for each test sample, including adversarial strengths/attacks not encountered during training. As a result, the framework dynamically selects defenses on a per-sample basis, generalizing to novel adversarial inputs and adapting to varying attack types and intensities. By moving beyond static defense mechanisms, this dynamic approach enhances resilience against previously unseen threats while maintaining computational efficiency, ensuring robust and practical performance.

\subsection{Final Evaluation Protocol}

During the final evaluation phase, each test sample is assigned to a corresponding defense model, ensuring that the selected strategy effectively mitigates the adversarial attack. 
To comprehensively assess \newframework{}'s effectiveness, we compute the Macro F1-Score for each selected defense model, offering a holistic measure of overall performance. This final metric is then compared with other baseline approaches, such as Oracle, random defense, the best static defense (state-of-the-art), and recommendation-based dynamic defense, highlighting \newframework{}'s robustness across diverse adversarial attack scenarios.


\section{Experimental Analysis}\label{sec:setup}

\subsection{Selected Datasets}

\begin{table}[]
\centering
\caption{Selected Intrusion Datasets}
\scalebox{0.95}{
\begin{tabular}{|c|c|c|c|c|}
\hline
\textbf{Dataset} & \textbf{Year} & \textbf{\begin{tabular}[c]{@{}c@{}}Number of \\ Features\end{tabular}} & \textbf{\begin{tabular}[c]{@{}c@{}}Number of \\ Attacks\end{tabular}} & \textbf{\begin{tabular}[c]{@{}c@{}}Number of \\ Samples\end{tabular}} \\ \hline
WUSTL-IIOT \cite{zolanvari2021wustl} & 2021 & 41 & 6  & 1M \\ \hline
UNSW-NB15 \cite{unsw15} & 2015 & 43 & 9  & 278K \\ \hline
X-IIoTID \cite{al2021x} & 2021 & 68 & 18  & 800K \\ \hline
Edge-IIoTID \cite{ferrag2022edge} & 2022 & 61 & 14  & 2.2M \\ \hline

\end{tabular}}
\label{datasets}
\end{table}

Table~\ref{datasets} summarizes the four intrusion datasets used in our paper. \textbf{WUSTL-IIoT}~\cite{zolanvari2021wustl} captures IIoT testbed traffic across realistic attack scenarios (41 features; 1M samples). \textbf{UNSW-NB15}~\cite{unsw15} mixes real and synthetic flows covering nine attack types (43 features; 278K samples). \textbf{X-IIoTID} \cite{al2021x} is a device- and connectivity-agnostic IIoT corpus with multi-view features from network, host, logs, and alerts (68 features; 800K instances). \textbf{Edge-IIoTset}~\cite{ferrag2022edge} spans cloud/fog/edge IoT traffic with 14 attack types (61 features; 2.2M instances). 

\subsection{Baselines}

To evaluate the effectiveness of \newframework{}, we compare its performance against multiple baseline approaches, enabling a systematic assessment of the benefits of dynamic defense selection. Each baseline serves as a reference point for quantifying \newframework{}'s improvements in robustness and efficiency under diverse adversarial attack scenarios.

\textbf{No Defense:} This baseline evaluates the performance of a standard Deep Neural Network (DNN) model under adversarial attacks without any defense. It helps us establish the lower performance bound, emphasizing the vulnerability of unprotected models to adversarial perturbations.

\textbf{Random Defense:} This baseline selects a defense for each test sample uniformly at random from the candidate portfolio, without any knowledge of the attack. To control for randomness, the evaluation is repeated over 100 independent runs, and the reported results correspond to the mean macro-F1 across runs, providing a stable estimate of expected performance while keeping computation tractable.

\textbf{Best Static (State-of-the-Art) Defense:} This baseline identifies the most effective defense model by evaluating all candidates on the attack training data. The selected model achieves the highest average performance across all adversarial attack types and is then applied to the attack test set. As a carefully tuned but non-adaptive approach, it represents the state-of-the-art ML-IDS defense. While it provides the optimal static configuration achievable via train-time selection, it cannot adapt to variations in test-time attacks, highlighting the limitations of non-dynamic strategies. As our best static defense baseline, we select, for each dataset, the most effective static defense identified through our own analysis. Specifically, we use \emph{PGD-Adversarial Training (PGD-AT)} for UNSW-NB15 \cite{madry2017towards}, \emph{TRADES} for WUSTL-IIoT \cite{zhang2019theoretically}, and the \emph{Gaussian Augmenter (GA)} for both X-IIoTID and Edge-IIoTset \cite{zantedeschi2017efficient}.












\textbf{Recommendation-Based Dynamic Defense:} The recommendation defense baseline uses a recommendation system to select the most appropriate defense model for each adversarial sample. This system measures the similarity between attack patterns and the training data using the Manhattan distance and identifies the optimal defense model based on this measure. The selected defense is then applied to the adversarial sample, providing a semi-informed approach to defense selection.

\textbf{Oracle Defense:} The Oracle represents the \emph{theoretical upper bound} of defense performance, achieved by selecting, for each test sample, the defense that performs best in hindsight. Such per-sample selection is infeasible in practice, as it would require evaluating all defenses at inference time. Comparing \newframework{}'s performance to the Oracle demonstrates how closely \newframework{} approximates this ideal, which is computationally infeasible for real-world applications.

\subsection{Experimental Setup}

\textbf{Hardware}. We conduct our experiments on a Linux virtual machine server equipped with a 16-core CPU, 32 GB of RAM, and an NVIDIA A100 GPU with 80 GB of memory.

\textbf{Evaluation Metric}. We select the Macro F1 score as our evaluation metric because it offers a balanced assessment of model performance across all classes, independent of class distribution. This metric is especially pertinent for datasets with imbalanced attack types, as it ensures that minority classes are appropriately represented in the evaluation.

\textbf{\newframework{} Performance Scoring}. The final defense performance is evaluated using a weighted scoring formula, which combines the number of samples handled by each defense model and its performance:

\begin{equation}
\text{Score} = \sum_{i=1}^{N} \left( \frac{\text{Sample Count}_i}{\text{Total Samples}} \times \text{Model Performance}_i \right)  
\end{equation}

where \( \text{Sample Count}_i \) represents the number of adversarial samples assigned to the \(i\)-th defense model, and \( \text{Model Performance}_i \) denotes the Macro F1 score of the \(i\)-th defense model. This formula calculates a weighted average of the defense models' performances, with the weight determined by the proportion of samples managed by each model. By doing so, it provides a holistic view of how well the dynamic algorithm selection performs across all assigned samples. A higher score reflects both the framework’s ability to assign the most suitable defense models and the overall effectiveness of those models in mitigating adversarial impacts.

\subsection{Dynamic Defense Selection Setup}
\label{defense-selection-setup}


\textbf{Model Selection:} 
We evaluate two ML models—XGBoost (XGB) and Random Forest (RF)—for the second-level learner. These models are chosen for their ability to effectively map attack patterns to optimal defense strategies. Each model is trained to predict the most suitable defense model for a given adversarial sample, and their performance is compared using the macro F1-score to identify the most effective approach for defense selection in IoT and IIoT environments.


\textbf{Unknown Adversarial Attacks Setup:}\label{sec:unseen-attacks-setup}
We evaluate defense selection under two types of ``unknowns'' adversarial conditions: (i) variations in attack strength, and (ii) previously unseen attacks (refer to Sec.~\ref{Dynamic_Defense_Selection_Algorithm} for details).

\paragraph{Variations in Attack Strength}\label{Perturbation-Strength Shifts} To probe generalization under intensity changes, we adopt an $\varepsilon$-shift protocol in which the performance matrix and optimal-defense labels are constructed on adversarial samples generated at a fixed training strength (e.g., $\varepsilon{=}0.1$), while evaluation is conducted at unseen strengths ($\varepsilon\in\{0.01,0.2,0.3\}$). All other components—data preprocessing, the defense portfolio, and EOAL-driven acquisition for the selector—remain unchanged. Macro-F1 is computed per attack model together with the cross-attack `Average', and results are compared against Oracle, best static, dynamic, random, and recommendation-based baselines.

\paragraph{Previously Unseen Attacks}\label{Structural gaps}
To emulate partial coverage of the threat landscape, we further consider training regimes in which one or more attack models are withheld: (i) exclude CaFA; (ii) exclude CaFA and AutoPGD; (iii) exclude CaFA, AutoPGD, and DF. The performance matrix and optimal-defense labels are rebuilt using only the remaining attacks. EOAL acquires labels from this reduced pool, and the selector is retrained accordingly. 

\textbf{Active Learning:} We restrict the proportion of labeled data used to train the second-level learner to a maximum of $50\%$ of the available adversarial training pool across all datasets. In practice, we examine an extremely low-label setting with only $1\%$ labeled data. For the ablation study, we systematically vary the cumulative label ratio at $1\%$, $10\%$, $20\%$, and $50\%$. Empirically, the best-performing label ratio (highest macro-F1) is dataset-specific: WUSTL-IIoT at \(1\%\), UNSW-NB15 at \(50\%\), Edge-IIoTest at \(10\%\), and X-IIoTID at \(50\%\).

\section{Results}\label{sec:results}

\begin{table*}[htbp]
\setlength{\tabcolsep}{4.5pt}
\centering
\caption{Final Performance (Macro F1-score) Comparison}
\begin{tabular}{|c|cccccc|cccccc|}
\hline
\textbf{(\%)}       & \multicolumn{1}{c|}{\textbf{Oracle}} & \multicolumn{1}{c|}{\textbf{\newframework{}}}  & \multicolumn{1}{c|}{\textbf{Dynamic}} & \multicolumn{1}{c|}{\textbf{Best Static}} & \multicolumn{1}{c|}{\textbf{Random}} & \textbf{No Defense} & \multicolumn{1}{c|}{\textbf{Oracle}} & \multicolumn{1}{c|}{\textbf{\newframework{}}}  & \multicolumn{1}{c|}{\textbf{Dynamic}} & \multicolumn{1}{c|}{\textbf{Best Static}} & \multicolumn{1}{c|}{\textbf{Random}} & \textbf{No Defense} \\ \hline
                    & \multicolumn{6}{c|}{UNSW-NB15}                                                                                                                                                                                              & \multicolumn{6}{c|}{WUSTL-IIoT}                                                                                                                                                                                             \\ \hline
\textbf{Clean Data} & \multicolumn{1}{c|}{99.12}           & \multicolumn{1}{c|}{94.79}          & \multicolumn{1}{c|}{96.82}            & \multicolumn{1}{c|}{88.65}                & \multicolumn{1}{c|}{89.37}           & 90.70               & \multicolumn{1}{c|}{99.51}           & \multicolumn{1}{c|}{98.84}          & \multicolumn{1}{c|}{98.84}            & \multicolumn{1}{c|}{94.81}                & \multicolumn{1}{c|}{89.08}           & 94.24               \\ \hline
\textbf{BIM}        & \multicolumn{1}{c|}{95.09}           & \multicolumn{1}{c|}{84.15}          & \multicolumn{1}{c|}{76.59}            & \multicolumn{1}{c|}{74.06}                & \multicolumn{1}{c|}{65.78}           & 30.78               & \multicolumn{1}{c|}{89.25}           & \multicolumn{1}{c|}{60.85}          & \multicolumn{1}{c|}{61.64}            & \multicolumn{1}{c|}{67.17}                & \multicolumn{1}{c|}{46.61}           & 28.89               \\ \hline
\textbf{FGSM}       & \multicolumn{1}{c|}{95.24}           & \multicolumn{1}{c|}{86.55}          & \multicolumn{1}{c|}{86.32}            & \multicolumn{1}{c|}{83.20}                & \multicolumn{1}{c|}{67.78}           & 43.20               & \multicolumn{1}{c|}{85.06}           & \multicolumn{1}{c|}{62.57}          & \multicolumn{1}{c|}{57.09}            & \multicolumn{1}{c|}{68.68}                & \multicolumn{1}{c|}{46.62}           & 28.83               \\ \hline
\textbf{PGD}        & \multicolumn{1}{c|}{95.09}           & \multicolumn{1}{c|}{84.15}          & \multicolumn{1}{c|}{76.59}            & \multicolumn{1}{c|}{74.06}                & \multicolumn{1}{c|}{65.78}           & 30.78               & \multicolumn{1}{c|}{89.25}           & \multicolumn{1}{c|}{60.85}          & \multicolumn{1}{c|}{61.64}            & \multicolumn{1}{c|}{67.17}                & \multicolumn{1}{c|}{46.61}           & 28.89               \\ \hline
\textbf{DF}         & \multicolumn{1}{c|}{71.76}           & \multicolumn{1}{c|}{62.40}          & \multicolumn{1}{c|}{59.76}            & \multicolumn{1}{c|}{36.97}                & \multicolumn{1}{c|}{20.84}           & 10.81               & \multicolumn{1}{c|}{64.13}           & \multicolumn{1}{c|}{40.47}          & \multicolumn{1}{c|}{28.64}            & \multicolumn{1}{c|}{2.13}                 & \multicolumn{1}{c|}{3.38}            & 1.41                \\ \hline
\textbf{AutoPGD}    & \multicolumn{1}{c|}{96.32}           & \multicolumn{1}{c|}{87.66}          & \multicolumn{1}{c|}{80.22}            & \multicolumn{1}{c|}{75.99}                & \multicolumn{1}{c|}{67.36}           & 29.52               & \multicolumn{1}{c|}{93.88}           & \multicolumn{1}{c|}{66.00}          & \multicolumn{1}{c|}{64.20}            & \multicolumn{1}{c|}{65.31}                & \multicolumn{1}{c|}{46.27}           & 31.01               \\ \hline
\textbf{ZOO}        & \multicolumn{1}{c|}{98.95}           & \multicolumn{1}{c|}{95.34}          & \multicolumn{1}{c|}{94.81}            & \multicolumn{1}{c|}{88.64}                & \multicolumn{1}{c|}{88.55}           & 83.51               & \multicolumn{1}{c|}{99.51}           & \multicolumn{1}{c|}{97.50}          & \multicolumn{1}{c|}{99.59}            & \multicolumn{1}{c|}{94.81}                & \multicolumn{1}{c|}{88.88}           & 90.47               \\ \hline
\textbf{CaFA}       & \multicolumn{1}{c|}{98.71}           & \multicolumn{1}{c|}{95.28}          & \multicolumn{1}{c|}{94.04}            & \multicolumn{1}{c|}{87.66}                & \multicolumn{1}{c|}{67.08}           & 28.71               & \multicolumn{1}{c|}{43.32}           & \multicolumn{1}{c|}{32.23}          & \multicolumn{1}{c|}{28.37}            & \multicolumn{1}{c|}{20.85}                & \multicolumn{1}{c|}{16.68}           & 3.07                \\ \hline
\textbf{SINIFGSM}   & \multicolumn{1}{c|}{94.23}           & \multicolumn{1}{c|}{83.97}          & \multicolumn{1}{c|}{82.51}            & \multicolumn{1}{c|}{80.55}                & \multicolumn{1}{c|}{65.56}           & 31.27               & \multicolumn{1}{c|}{90.22}           & \multicolumn{1}{c|}{61.49}          & \multicolumn{1}{c|}{71.04}            & \multicolumn{1}{c|}{67.63}                & \multicolumn{1}{c|}{44.43}           & 30.45               \\ \hline
\textbf{VNIFGSM}    & \multicolumn{1}{c|}{95.19}           & \multicolumn{1}{c|}{85.80}          & \multicolumn{1}{c|}{87.85}            & \multicolumn{1}{c|}{84.89}                & \multicolumn{1}{c|}{69.04}           & 33.56               & \multicolumn{1}{c|}{95.34}           & \multicolumn{1}{c|}{70.87}          & \multicolumn{1}{c|}{74.26}            & \multicolumn{1}{c|}{81.67}                & \multicolumn{1}{c|}{49.81}           & 28.68               \\ \hline
\textbf{Average}    & \multicolumn{1}{c|}{93.40}           & \multicolumn{1}{c|}{\textbf{85.03}} & \multicolumn{1}{c|}{{\ul 82.08}}      & \multicolumn{1}{c|}{76.23}                & \multicolumn{1}{c|}{64.20}           & 35.79               & \multicolumn{1}{c|}{83.33}           & \multicolumn{1}{c|}{\textbf{61.43}} & \multicolumn{1}{c|}{{\ul 60.72}}            & \multicolumn{1}{c|}{59.49}          & \multicolumn{1}{c|}{43.25}           & 30.19               \\ \hline
                    & \multicolumn{6}{c|}{X-IIoTID}                                                                                                                                                                                               & \multicolumn{6}{c|}{Edge-IIoTest}                                                                                                                                                                                           \\ \hline
\textbf{Clean Data} & \multicolumn{1}{c|}{98.71}           & \multicolumn{1}{c|}{98.17}          & \multicolumn{1}{c|}{97.88}            & \multicolumn{1}{c|}{93.11}                & \multicolumn{1}{c|}{96.98}           & 98.40               & \multicolumn{1}{c|}{100.00}          & \multicolumn{1}{c|}{97.80}          & \multicolumn{1}{c|}{100.00}           & \multicolumn{1}{c|}{94.45}                & \multicolumn{1}{c|}{91.24}           & 94.41               \\ \hline
\textbf{BIM}        & \multicolumn{1}{c|}{95.91}           & \multicolumn{1}{c|}{80.48}          & \multicolumn{1}{c|}{82.20}            & \multicolumn{1}{c|}{77.77}                & \multicolumn{1}{c|}{55.59}           & 10.80               & \multicolumn{1}{c|}{98.87}           & \multicolumn{1}{c|}{94.80}          & \multicolumn{1}{c|}{91.80}            & \multicolumn{1}{c|}{94.18}                & \multicolumn{1}{c|}{79.34}           & 84.29               \\ \hline
\textbf{FGSM}       & \multicolumn{1}{c|}{94.81}           & \multicolumn{1}{c|}{78.65}          & \multicolumn{1}{c|}{71.96}            & \multicolumn{1}{c|}{75.87}                & \multicolumn{1}{c|}{54.20}           & 22.77               & \multicolumn{1}{c|}{99.17}           & \multicolumn{1}{c|}{93.47}          & \multicolumn{1}{c|}{93.28}            & \multicolumn{1}{c|}{94.40}                & \multicolumn{1}{c|}{80.62}           & 85.55               \\ \hline
\textbf{PGD}        & \multicolumn{1}{c|}{95.91}           & \multicolumn{1}{c|}{80.48}          & \multicolumn{1}{c|}{82.20}            & \multicolumn{1}{c|}{77.77}                & \multicolumn{1}{c|}{55.59}           & 10.80               & \multicolumn{1}{c|}{98.87}           & \multicolumn{1}{c|}{94.80}          & \multicolumn{1}{c|}{91.80}            & \multicolumn{1}{c|}{94.18}                & \multicolumn{1}{c|}{79.34}           & 84.29               \\ \hline
\textbf{DF}         & \multicolumn{1}{c|}{99.16}           & \multicolumn{1}{c|}{97.53}          & \multicolumn{1}{c|}{97.74}            & \multicolumn{1}{c|}{89.47}                & \multicolumn{1}{c|}{54.14}           & 1.79                & \multicolumn{1}{c|}{10.44}           & \multicolumn{1}{c|}{5.90}           & \multicolumn{1}{c|}{5.61}             & \multicolumn{1}{c|}{1.50}                 & \multicolumn{1}{c|}{2.16}            & 1.54                \\ \hline
\textbf{AutoPGD}    & \multicolumn{1}{c|}{95.87}           & \multicolumn{1}{c|}{81.41}          & \multicolumn{1}{c|}{83.80}            & \multicolumn{1}{c|}{79.11}                & \multicolumn{1}{c|}{56.69}           & 10.22               & \multicolumn{1}{c|}{99.79}           & \multicolumn{1}{c|}{97.94}          & \multicolumn{1}{c|}{97.30}            & \multicolumn{1}{c|}{94.30}                & \multicolumn{1}{c|}{85.30}           & 84.11               \\ \hline
\textbf{ZOO}        & \multicolumn{1}{c|}{98.35}           & \multicolumn{1}{c|}{98.14}          & \multicolumn{1}{c|}{97.83}            & \multicolumn{1}{c|}{93.11}                & \multicolumn{1}{c|}{94.79}           & 82.53               & \multicolumn{1}{c|}{100.00}          & \multicolumn{1}{c|}{98.59}          & \multicolumn{1}{c|}{100.00}           & \multicolumn{1}{c|}{94.56}                & \multicolumn{1}{c|}{91.29}           & 94.41               \\ \hline
\textbf{CaFA}       & \multicolumn{1}{c|}{89.33}           & \multicolumn{1}{c|}{71.34}          & \multicolumn{1}{c|}{72.24}            & \multicolumn{1}{c|}{68.58}                & \multicolumn{1}{c|}{54.29}           & 1.51                & \multicolumn{1}{c|}{72.61}           & \multicolumn{1}{c|}{70.51}          & \multicolumn{1}{c|}{72.36}            & \multicolumn{1}{c|}{55.20}                & \multicolumn{1}{c|}{41.57}           & 29.97               \\ \hline
\textbf{SINIFGSM}   & \multicolumn{1}{c|}{97.06}           & \multicolumn{1}{c|}{90.15}          & \multicolumn{1}{c|}{82.43}            & \multicolumn{1}{c|}{77.19}                & \multicolumn{1}{c|}{57.28}           & 12.73               & \multicolumn{1}{c|}{98.90}           & \multicolumn{1}{c|}{92.70}          & \multicolumn{1}{c|}{92.19}            & \multicolumn{1}{c|}{94.20}                & \multicolumn{1}{c|}{79.59}           & 84.29               \\ \hline
\textbf{VNIFGSM}    & \multicolumn{1}{c|}{96.29}           & \multicolumn{1}{c|}{85.46}          & \multicolumn{1}{c|}{85.08}            & \multicolumn{1}{c|}{83.34}                & \multicolumn{1}{c|}{55.72}           & 8.93                & \multicolumn{1}{c|}{99.38}           & \multicolumn{1}{c|}{94.80}          & \multicolumn{1}{c|}{94.41}            & \multicolumn{1}{c|}{94.32}                & \multicolumn{1}{c|}{83.31}           & 86.59               \\ \hline
\textbf{Average}    & \multicolumn{1}{c|}{95.85}           & \multicolumn{1}{c|}{\textbf{84.85}} & \multicolumn{1}{c|}{{\ul 83.94}}      & \multicolumn{1}{c|}{80.25}                & \multicolumn{1}{c|}{59.81}           & 18.01               & \multicolumn{1}{c|}{86.45}           & \multicolumn{1}{c|}{\textbf{82.61}} & \multicolumn{1}{c|}{{\ul 82.08}}      & \multicolumn{1}{c|}{79.65}                & \multicolumn{1}{c|}{69.17}           & 70.56               \\ \hline
\end{tabular}
\label{model_Performance}
\end{table*}

\subsection{\newframework{} Defense Performance}
Table~\ref{model_Performance} presents a comparative analysis of \newframework{} and the selected baselines (Oracle, dynamic, best static, random, and no defense) under the scenario where the perturbation amount is varied for the same adversarial attack types, i.e., \textit{variations in attack strength}. The table reports the macro F1 score for each adversarial attack, as well as the average score across all perturbation levels and attack scenarios. It is evident that \newframework{} substantially outperforms the dynamic, best static, random, and no defense baselines, achieving performance nearly equivalent to that of the Oracle. Table \ref{improvement_rate} presents both the maximum and average performance improvements of \newframework{} compared to these baselines. These results highlight \newframework{}'s effectiveness in maintaining robustness as the strength of adversarial perturbations changes.

\textbf{Comparison with Random and Best Static Defenses:}
\newframework{} exhibits the most significant improvement in performance for the DF attack among all considered adversarial attacks. It achieves substantial improvements over both random and best static defenses, with performance gains of 199.4\% and 68.8\% on the UNSW-NB15 dataset, 1097.3\% and 1800.1\% on the WUSTL-IIoT dataset, 80.2\% and 16.8\% on the X-IIoTID dataset, and 173.3\% and 293.4\% on the Edge-IIoTest dataset, respectively, for the DF attack. This demonstrates that, especially in the case of stronger attacks, \newframework{} substantially outperforms random and best static, highlighting its effectiveness and reliability in optimizing defense model allocation. As shown in Table \ref{model_Performance}, static defenses are more vulnerable to white-box attacks, as attackers can exploit their weaknesses, whereas in black-box attacks like ZOO, static defenses maintain relatively stable performance. However, \newframework{} still achieves a notable performance gain, improving `Average' up to 12\% over the best static defense. Our method also shows average improvements over both baselines, highlighting its effectiveness and adaptability in dynamically assigning optimal defense strategies.

\textbf{Comparison with Dynamic Defense:}
When compared to the recommendation-based dynamic defense strategy, Table~\ref{model_Performance} highlights \newframework{}'s superior robustness and adaptability across a variety of adversarial scenarios and datasets. For instance, \newframework{} achieves up to 41.3\% improvement in macro F1 scores, as seen in DF (40.47\% vs. 28.64\%) on WUSTL-IIoT. Unlike dynamic defenses, which rely on distance to adjust model parameters, \newframework{} employs an adaptive mechanism that dynamically mitigates perturbations without dependence on distance-based adjustments. This approach enables \newframework{} to avoid overfitting to specific attack patterns, enhancing its resilience against a wide range of adversarial threats. Another key aspect of \newframework{}'s superiority is its balanced performance across both white-box and black-box attacks. Under the black-box ZOO attack, \newframework{} maintains high F1 scores (e.g., 95.34\% on UNSW-NB15 and 98.14\% on X-IIoTID), closely matching or exceeding dynamic defenses (94.81\% and 97.83\%, respectively). These results highlight that \newframework{} achieves a superior trade-off, significantly enhancing robustness in the challenging white-box setting while remaining competitive against black-box threats. As a result, as shown in Table \ref{improvement_rate}, \newframework{}'s average F1 scores reflect improvements of up to 5.0\% over dynamic defenses across all attack scenarios.

\begin{table}[]
\centering
\caption{\newframework{} F1-score Improvement Rate}
\begin{tabular}{|cc|c|c|c|}
\hline
\multicolumn{2}{|c|}{Improvement Rate (\%)}                   & Dynamic & Best Static & Random  \\ \hline
\multicolumn{1}{|c|}{\multirow{2}{*}{UNSW-NB15}}    & Max     & 9.9    & \underline{68.8}       & \textbf{199.4}  \\ \cline{2-5} 
\multicolumn{1}{|c|}{}                              & Average & 3.9    & \underline{15.2}       & \textbf{46.1}   \\ \hline
\multicolumn{1}{|c|}{\multirow{2}{*}{WUSTL-IIoT}}   & Max     & 41.3   & \textbf{1800.1}     & \underline{1097.3} \\ \cline{2-5} 
\multicolumn{1}{|c|}{}                              & Average & 5.0    & \textbf{201.0}      & \underline{157.7}  \\ \hline
\multicolumn{1}{|c|}{\multirow{2}{*}{X-IIoTID}}     & Max     & 9.4    & \underline{16.8}       & \textbf{80.2}   \\ \cline{2-5} 
\multicolumn{1}{|c|}{}                              & Average & 1.2    & \underline{5.7}        & \textbf{44.9}   \\ \hline
\multicolumn{1}{|c|}{\multirow{2}{*}{Edge-IIoTest}} & Max     & 5.2    & \textbf{293.4}      & \underline{173.3}  \\ \cline{2-5} 
\multicolumn{1}{|c|}{}                              & Average & 1.1    & \underline{36.5}       & \textbf{39.0}   \\ \hline
\end{tabular}
\label{improvement_rate}
\end{table}


\begin{figure}[]
\centering\includegraphics[scale=0.35]{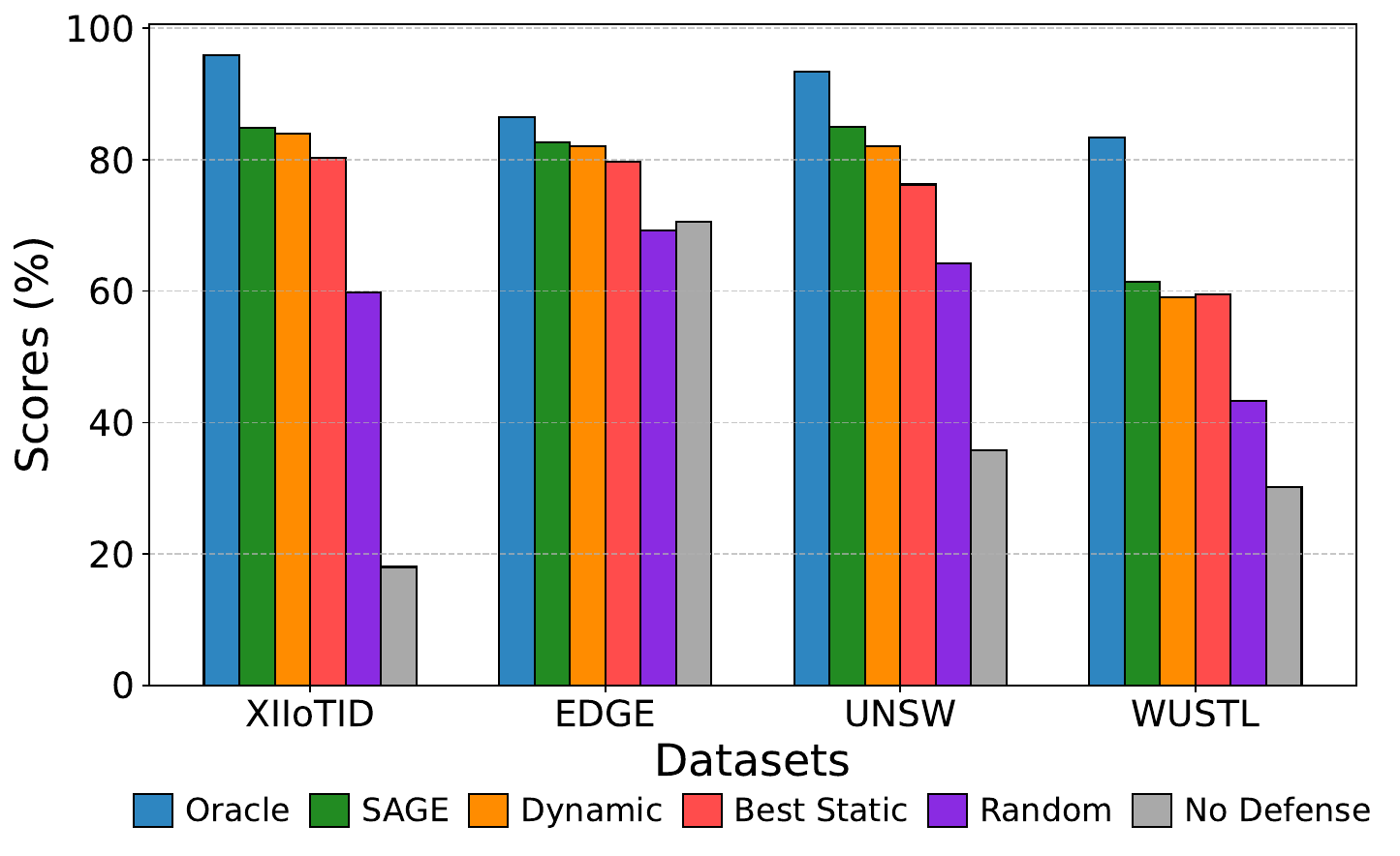}
\caption{Average performance (Macro-F1) across the five baselines}
\label{Average_Main_Performance}
\end{figure}

\textbf{Comparison with Oracle: }
\newframework{} demonstrates a remarkably small performance gap with the Oracle, underscoring its effectiveness in dynamically selecting defenses across diverse adversarial conditions. In black-box attacks like ZOO, where it achieves 95.34\% vs. 98.95\% on UNSW-NB15, 97.50\% vs. 99.51\% on WUSTL-IIoT, and 98.14\% vs. 98.35\% on X-IIoTID, demonstrating near-Oracle performance without requiring prior knowledge of attack mechanisms. Even in the most challenging case, the DF attack, \newframework{}'s 97.53\% closely rivals the Oracle's 99.16\%, suggesting that its adaptive modeling enables it to effectively resist the smallest perturbation attacks. These results highlight \newframework{}'s ability to allocate defenses effectively without requiring exhaustive model evaluations, making it both efficient and overhead. On average, the performance difference remains minimal, with a 3.84\% gap in Edge-IIoTest and 8.1\% in UNSW-NB15, reinforcing \newframework{}'s adaptability even in more complex adversarial scenarios.

\textbf{Clean Data Performance:}
Table~\ref{model_Performance} presents a comparison of clean data performance for \newframework{} against baseline methods. The results are evaluated using macro F1 scores under clean (no adversarial attack) conditions. Specifically, \newframework{} achieves F1 scores of 94.79\% on UNSW-NB15, 98.84\% on WUSTL-IIoT, 98.17\% on X-IIoTID, and 97.80\% on Edge-IIoTest, closely approaching the Oracle's near-optimal scores of 99.12\%, 99.51\%, 98.71\%, and 100.00\%, respectively. Compared to Dynamic and No Defense baselines, \newframework{} consistently delivers superior or comparable performance, while significantly outperforming Random and Best Static in most cases. As illustrated in Table~\ref{model_Performance} (Clean Data), \newframework{} maintains top-tier performance on clean data, ranking at the first or second level relative to all baselines beside the Oracle. These results highlight that \newframework{} achieves enhanced robustness without compromising performance under benign conditions, underscoring its effectiveness across diverse datasets.

\textbf{Insights:} The results reflect \newframework{}'s ability to approach the theoretical upper bound established by the Oracle while demonstrating its flexibility in handling diverse adversarial scenarios. Since the Oracle determines the best performance by testing every dataset across all models, achieving this level of optimality in practice would require significant effort and resources, making it impractical for real-world applications. In contrast, the \newframework{} dynamically allocates defense strategies using a machine learning-based approach. This method eliminates the need for manually selecting the optimal defense for each attack, showcasing the \newframework{}'s adaptability in addressing complex adversarial scenarios. Moreover, \newframework{}'s effectiveness lies in its robust generalization across both white-box and black-box attack scenarios, a flexibility that the Oracle's idealized approach cannot replicate in practical settings. For instance, \newframework{}'s performance in complex attacks like CaFA and DF, highlights its ability to mitigate sophisticated perturbations effectively, even without the Oracle's perfect knowledge. This adaptability stems from \newframework{}'s integrated modeling of spatial and temporal dependencies, which allows it to capture nuanced patterns in data distributions, unlike the Oracle's reliance on exhaustive testing.

\begin{table}[htbp]
\centering
\caption{Unseen Adversarial Attack Performance (Macro F1, Averages)}
\resizebox{\columnwidth}{!}{%
\begin{tabular}{|c|c|c|c|c|}
\hline
\textbf{Average (\%)} & \textbf{UNSW-NB15} & \textbf{WUSTL-IIoT} & \textbf{X-IIoTID} & \textbf{Edge-IIoTset} \\ \hline
No Exclusion & \textbf{85.03} & \underline{61.43} & \textbf{84.85} & \textbf{82.61} \\ \hline
\begin{tabular}[c]{@{}c@{}}Exclude\\ CaFA\end{tabular} & \underline{84.08} & 57.26 & \underline{83.74} & 79.72 \\ \hline
\begin{tabular}[c]{@{}c@{}}Exclude\\ CaFA\&AutoPGD\end{tabular} & 83.99 & \textbf{61.71} & 83.25 & 77.98 \\ \hline
\begin{tabular}[c]{@{}c@{}}Exclude\\ CaFA\&AutoPGD\&DF\end{tabular} & 79.50 & 57.78 & 80.85 & \underline{81.40} \\ \hline
\end{tabular}
}
\label{unseen_attack_Performance}
\end{table}

\subsection{Unseen Adversarial Attacks Performance}

To evaluate robustness to unknown attack models, i.e., \textit{previously unseen attack types}, we remove one or more attack models from the training set and then evaluate on the full suite, as detailed in Sec.~\ref{sec:unseen-attacks-setup}. Table~\ref{unseen_attack_Performance} reports `Average' macro-F1. The four evaluation conditions are: no exclusion (all attacks included); exclude CaFA; exclude CaFA and AutoPGD; exclude CaFA, AutoPGD, and DF.


First, we analyze individual attack performances on a per-attack basis. Withholding a given attack model primarily affects the withheld model while leaving other models largely unaffected. For instance, on WUSTL-IIoT, DF decreases from $36.24\%$ (when DF is included in training) to $0.99\%$ when DF is withheld, corresponding to a drop of $35.25$ points; on Edge-IIoTset, CaFA declines from $70.51\%$ to $31.13\%$ when CaFA is withheld, a reduction of $39.38$ points. By contrast, ZOO remains near ceiling across conditions, with performance ranging from $97\%$ to $99\%$ on all four datasets. This pattern suggests that the selector establishes distinct decision regions for each attack model: removing a model leaves its decision region under-represented without causing systemic degradation in the performance of other models.


Second, despite such targeted drops, the system-level ‘Average’ remains comparatively stable across exclusion regimes, evidencing cross-attack generalization. From ‘No Exclusion’ to ‘Exclude CaFA+AutoPGD+DF’, the average decreases by only 5.53 points on UNSW-NB15 (85.03\% to 79.50\%), 3.65 points on WUSTL-IIoT (61.43\% to 57.78\%), 4.00 points on X-IIoTID (84.85\% to 80.85\%), and 1.21 points on Edge-IIoTest (82.61\% to 81.40\%). Clean traffic also stays high across conditions (e.g., UNSW 94.79--96.03\%, Edge 97.68--98.69\%), reinforcing that robustness is preserved at the system level even when specific attacks are unknown at training time.


Overall, the \newframework{} exhibits localized sensitivity (strong drops on the held-out model) and global stability (modest change in the cross-attack average), which together characterize reliable defense selection under distribution shift.

\begin{figure}[]
\centering\includegraphics[scale=0.35]{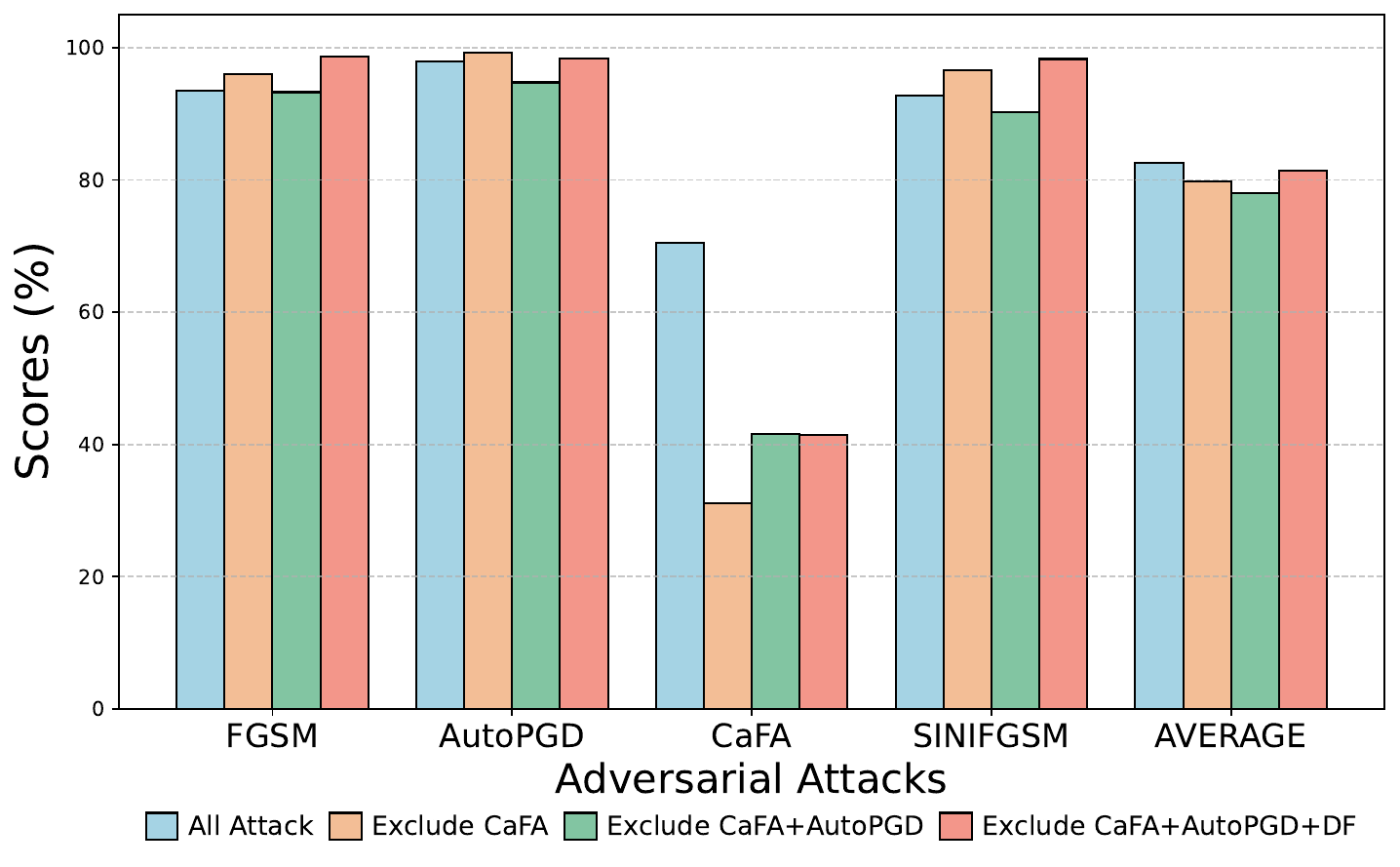}
\caption{Edge-IIoTest Unseen Adversarial Attack Performance Comparison}
\label{SAGE_Exclude_Attack}
\end{figure}


\subsection{Overhead Analysis}

The overhead analysis evaluates the computational efficiency of \newframework{} by comparing the per-sample defense selection time against Oracle, Dynamic, and Best-static baselines. Table~\ref{processing_time} reports the processing time in ms/sample, illustrating the substantial efficiency advantage of \newframework{} over the competing methods. \newframework{} maintains low inference overhead, requiring less than $1.01$~ms per sample across datasets. By contrast, the Oracle requires approximately $24$--$27$~ms per sample, making \newframework{} up to $29\times$ faster. The efficiency gain comes from executing a single learned selector and dispatching one defense per input, rather than exhaustively evaluating the entire defense portfolio as the Oracle does. These dramatic reductions highlight \newframework{}’s active learning and machine learning-driven adaptive mechanism that enables rapid defense selection without the resource-intensive enumeration required by the Oracle, making it far more practical for real-world applications where computational efficiency is critical.

Compared to the Dynamic baseline, which leverages distance for real-time parameter adjustments, \newframework{} exhibits slightly higher processing times, with a gap of up to 0.21 ms/sample across the datasets. However, this marginal increase is offset by \newframework{}’s superior robustness, as demonstrated in prior sections, particularly against gradient-based attacks (e.g., up to 41\% F1 score improvement for one attack and up to 5\% on average). The Dynamic approach, while computationally efficient, often sacrifices generalization due to its reliance on real-time distance calculations, which can lead to suboptimal defense strategies in complex scenarios. 

Against the Best Static baseline, \newframework{} offers significant efficiency gains, achieving speedups of $2.48\times$ (0.89 vs. 2.21 ms/sample) on UNSW-NB15, $2.64\times$ (0.92 vs. 2.43 ms/sample) on WUSTL-IIoT, $2.51\times$ (0.85 vs. 2.13 ms/sample) on X-IIoTID, and $2.40\times$ (1.01 vs. 2.42 ms/sample) on Edge-IIoTest. While Best Static is more efficient than the Oracle, it still lacks the adaptability of \newframework{}, underscoring \newframework{}'s efficiency in accelerating defense selection while maintaining strong performance.

\begin{table}[htbp]
\setlength{\tabcolsep}{4pt}
\centering
\caption{Processing Time (ms/Sample) Comparison}
\label{processing_time}
\begin{tabular}{lcccc}
\toprule
\textbf{ms/Sample}   & \textbf{UNSW-NB15} & \textbf{WUSTL-IIoT} & \textbf{X-IIoTID} & \textbf{Edge-IIoTest} \\ 
\midrule
\newframework{}              & \underline{0.89}              & \underline{0.92}                & \underline{0.85}              & \underline{1.01}                 \\ 
Oracle            & 24.28             & 26.70               & 23.44             & 26.58                \\ 
Dynamic           & \textbf{0.68}              & \textbf{0.69}                & \textbf{0.63}              & \textbf{0.72}                 \\ 
Best Static       & 2.21              & 2.43                & 2.13              & 2.42                 \\ 
\bottomrule
\end{tabular}
\end{table}

\subsection{Ablation Study}
\label{sec:ablation}

\subsubsection{Active Learning Methods}

\begin{figure}[]
\centering\includegraphics[scale=0.35]{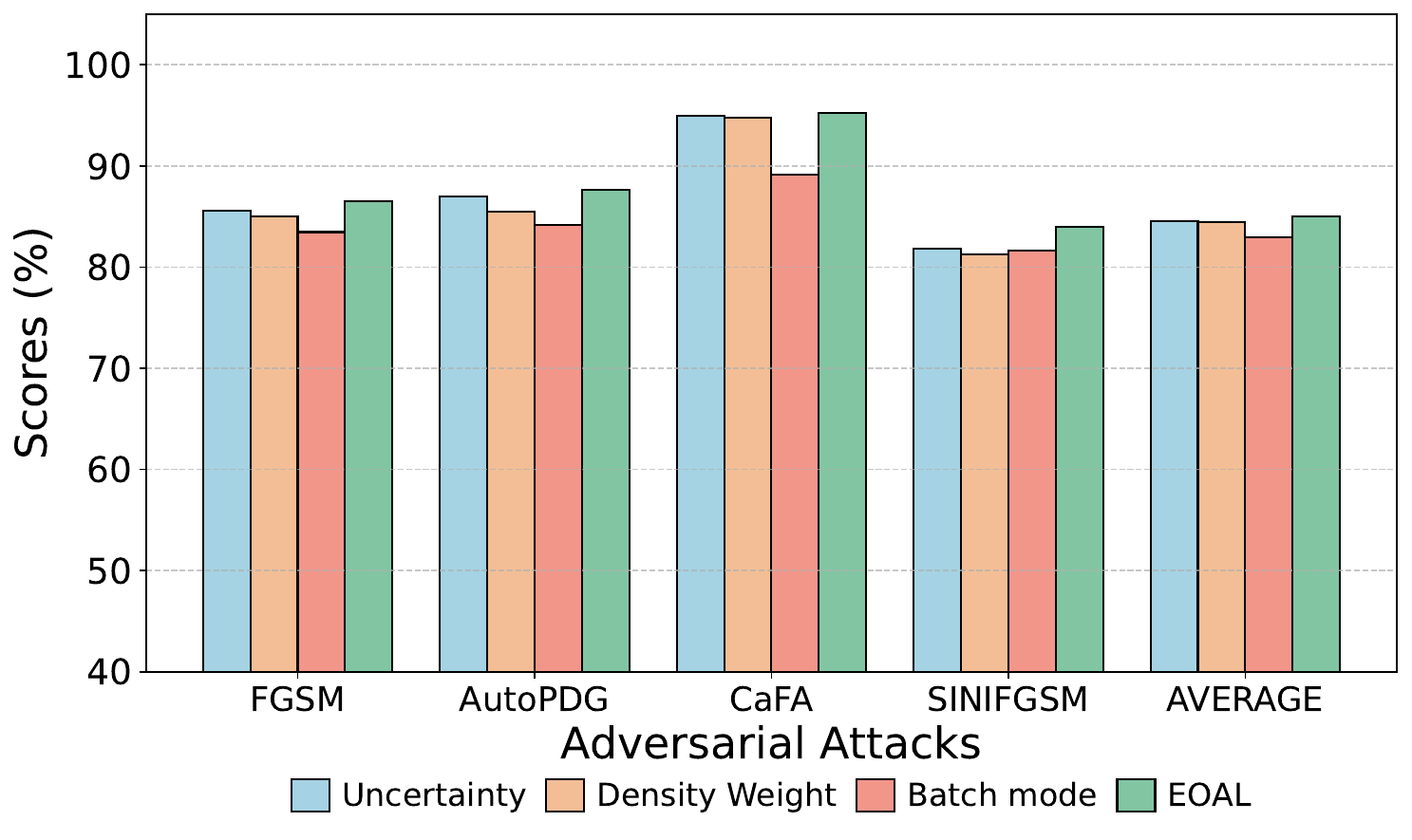}
\caption{Active Learning Performance Comparison (UNSW-NB15)}
\label{ActiveLearning_Strategies_Performance}
\end{figure}

We evaluate several active learning (AL) sampling strategies for their effect on the robustness of the second-level defense selector, including conventional uncertainty sampling, density weight sampling, batch mode sampling~\cite{ren2021survey}, and Entropic Open-set Active Learning (EOAL). As shown in Fig.~\ref{ActiveLearning_Strategies_Performance}, across representative white-box gradient attacks, e.g., FGSM, AutoPGD, and composite attacks, e.g., CaFA, EOAL attains the best or tied-best macro-F1 on most datasets and exhibits reduced variance across attacks, indicating superior stability and generalization. Three empirical insights emerge from our evaluation. First, EOAL uses entropy-based acquisition with explicit open-set modeling to consistently query the most informative samples, focusing annotation on decision regions associated with novel or rare attack behaviors and avoiding adversarial hard-but-uninformative cases. Second, its batch-quota acquisition under comparable labeling budgets (50\%, 20\%, 10\%, 1\%) preserves diversity and limits redundancy, yielding higher information per label and pushing the selector toward near full-data performance at substantially lower annotation rates. Third, under our $\varepsilon$-shifted protocol---training with $\varepsilon=0.1$ and testing on unseen $\varepsilon \in \{0.01, 0.2, 0.3\}$---EOAL maintains a stable mapping from inputs to optimal defenses, demonstrating robustness to distribution shift. Together with the comparative results, these insights justify our adoption of EOAL as the active learning component of \newframework{}.

\subsubsection{Different Proportion of Training Data}
\begin{figure}[]
\centering\includegraphics[scale=0.35]{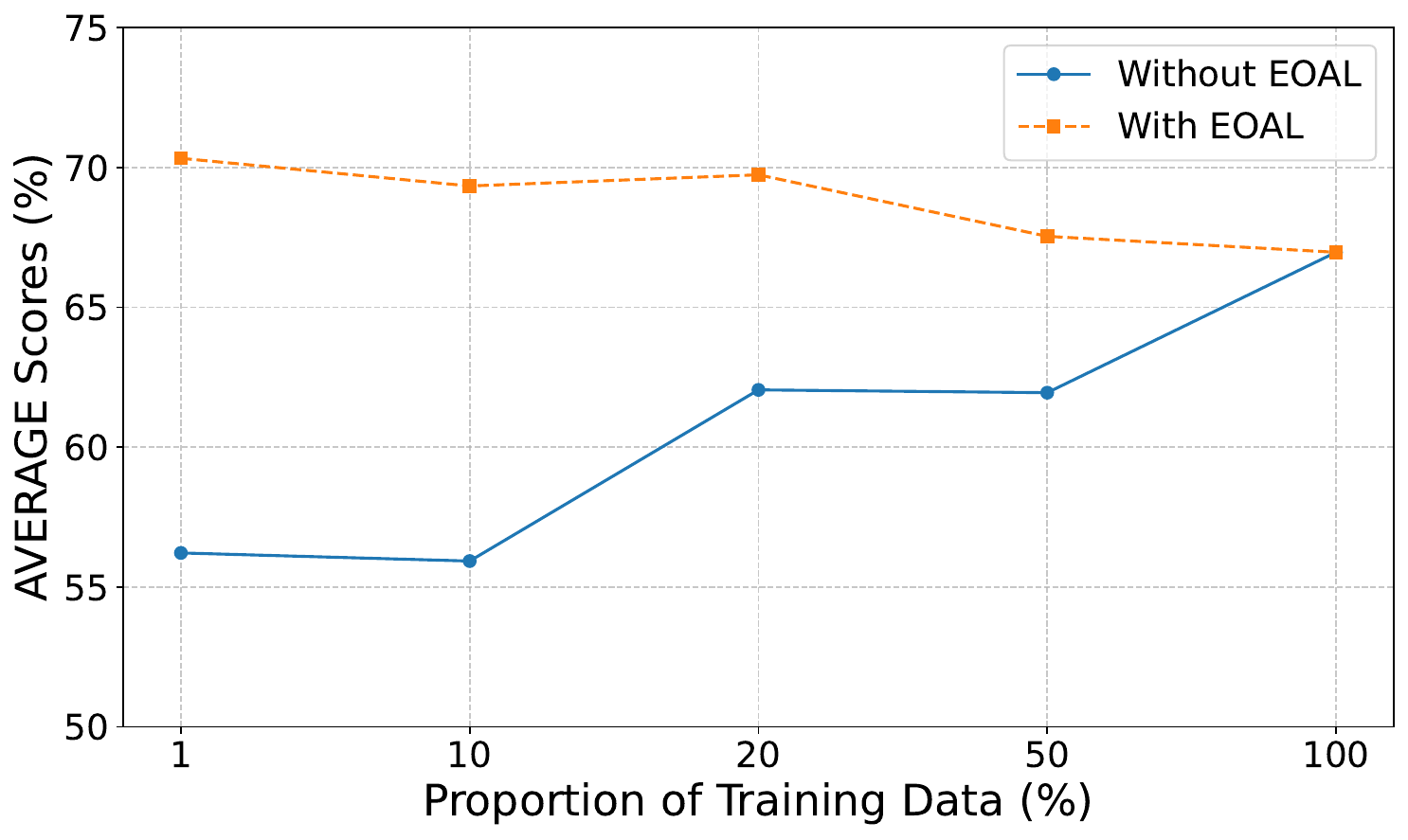}
\caption{Different Proportion of Training Data Comparison (WUSTL-IIoT)}
\label{WUSTL_Diff_Prop_Training_Data_Average}
\end{figure}

We examine SAGE’s label-efficiency under EOAL by varying the proportion of labeled attack training data. As shown in Fig.~\ref{WUSTL_Diff_Prop_Training_Data_Average}, macro-F1 increases with larger budgets, yet under EOAL the curve is steep at low label fractions: strong performance already emerges at \(1\%\) of the pool, approaching the performance achieved with substantially larger budgets (e.g., \(10\%\text{–}50\%\)). By explicitly targeting sparsely covered regions of the attack space, EOAL can match—or even exceed—full-label training with only \(1\%\) supervision, delivering gains of up to 3.4 points. By contrast, without EOAL, using stratified random sampling, performance improves more slowly and requires substantially larger labeled fractions to reach comparable results. This advantage stems from EOAL’s entropy-ranked, diversity-aware acquisition, which concentrates annotation on the most informative regions and reduces redundancy; as a result, the second-level learner approaches its full-data quality with a small fraction of labels, cutting annotation cost and training time while preserving robustness.


\section{Conclusion}\label{sec:conclusion}

In high-stakes operational networks, ML-based intrusion detection systems (ML-IDS) must contend with evolving attack geometries and distribution shifts. To address this, our framework, \newframework{}, formulates adversarially robust ML-IDS as a per-sample defense selection problem. It couples a curated defense portfolio with an EOAL-trained selector, enabling the system to choose the most effective defense for each input at inference while maintaining label- and compute-efficiency and preserving accuracy on clean data. \newframework{} achieves near-Oracle performance, attaining a best-case macro-F1 within 3.8\% of the Oracle while accelerating per-sample computation by up to $29\times$. It delivers the largest gains over baselines, with up to a 5.0\% improvement relative to a recommendation-based dynamic baseline, and remains robust under unseen attacks, with the cross-attack average decreasing by only 1.21 points.

\section*{Acknowledgments}
This work has been funded in part by NSF, with award numbers \#1826967, \#1911095, \#2003279, \#2052809, \#2100237, \#2112167, \#2112665, and in part by PRISM and CoCoSys, centers in JUMP 2.0, an SRC program sponsored by DARPA.


\ifCLASSOPTIONcaptionsoff
  \newpage
\fi



\bibliographystyle{IEEEtran}
%
\bibliography{biblio}




%








\end{document}